# The Complexity of Molecular Interactions and Bindings between Cyclic Peptide and Inhibit Polymerase A and B1 (PAC-PB1N) H1N1

Arli Aditya Parikesit[a], Harry Noviardi[a], Djati Kerami[b], Usman Sumo Friend Tambunan[a]

**ABSTRACT**

The Influenza/H1N1 virus has caused hazard in the public health of many countries. Hence, existing influenza drugs could not cope with H1N1 infection due to the high mutation rate of the virus. In this respect, new method to block the virus was devised. The polymerase PAC-PB1N enzyme is responsible for the replication of H1N1 virus. Thus, novel inhibitors were developed to ward off the functionality of the enzyme. In this research, cyclic peptides has been chosen to inhibit PAC-PB1N due to its proven stability in reaching the drug target. thus, computational method for elucidating the molecular interaction between cyclic peptides and PAC-PB1N has been developed by using the LigX tools from MOE 2008.10 software. The tools could render the bindings that involved in the interactions. The interactions between individual amino acid in the inhibitor and enzyme could be seen as well. Thus, the peptide sequences of CKTTC and CKKTC were chosen as the lead compounds. In this end, the feasibility of cyclic peptides to act as drug candidate for H1N1 could be exposed by the 2D and 3D modeling of the molecular interactions.

[a] Bioinformatics Research Group, Department of Chemistry, Faculty of Mathematics and Science, University of Indonesia, Depok 16424, Indonesia

[b] Mathematics Computation Group, Department of Mathematics, Faculty of Mathematics and Science, University of Indonesia, Depok 16424, Indonesia

**Corresponding author e-mail address: usman@ui.ac.id**

## Introduction

WHO has declared that Influenza A/H1N1 as serious treat in the public health (WHO, 2014). Therefore, drugs must be developed to overcome the infection. Currently, scientist is developing H1N1 drugs based on molecular biology information system (Chang, Huang, & Chen, 2011). Influenza A H1N1 virus carried eight negative-strand RNA genome segments, and each one was bound by the viral encoded RNA-dependent RNA polymerase (RdRp) complex (Larsen et al., 2014).The heterotrimeric complex of PA, PB1, and PB2 are working together as RNA polymerase enzyme that responsible for the reproduction of Viral RNA (Y. Liu, Lou, Bartlam, & Rao, 2009).

The influenza virus polymerase can replicate de novo its viral genome RNA through two steps. The replication step involves synthesis of complementary RNA (cRNA) from a viral RNA (vRNA) template, followed by synthesis of vRNA from cRNA to complete the genomic RNA replication process (Y. Liu et al., 2009). The transcription step involves viral mRNA transcription from vRNA by snatching capped primers from nascent host cell mRNA (Y. Liu et al., 2009).

In previous research, viral RNA synthesis could be blocked by the specific inhibition of viral polymerase complex formation by using small peptide that bound to the protein-protein interaction domain that responsible for hetero-oligomerization amongst the individual subunits. The small peptide was consisted of 25 amino acid corresponding to the PA-binding domain of PB1 that inhibits the polymerase activity of influenza A virus and obstruct the viral spread (Ghanem et al., 2007). Thus, the peptide sequence nomenclature is always following the one-letter system from IUPAC-IUB (IUPAC-IUB, 1984). In one hand, Ribavirin was also tested as H1N1 drugs in animals (in vivo) and in the clinical trials (Rowe et al., 2010; Wilson et al., 1984). In the other hand, the determination of peptide sequence PTLLFL was indispensable for warding off influenza A virus (Obayashi, Yoshida, & Park, 2008). The formation of hydrogen bonds of the PTLLFL peptide sequence with PBA were conducted by the last two residues (Obayashi et al., 2008). Thus, the cyclic peptide has proven as stable and robust drug candidate (Bogdanowich-Knipp, Chakrabarti, Williams, Dillman, & Siahaan, 1999; Jois, Tambunan, Chakrabarti, & Siahaan, 1996).Therefore, the objective of this research is to develop cyclic peptide-based drug to overcome the H1N1 infection.

## Methodology

The pipeline was prepared in accordance to our methods (A. A. Parikesit, Kinanty, & Tambunan, 2013; A. Parikesit, 2010; Tambunan, Parikesit, Dephinto, & Sipahutar, 2014; Tambunan, Zahroh, Utomo, & Parikesit, 2014). The Polymerase A and B1 sequences were retrieved from NCBI database (NCBI, 2013; Wheeler et al., 2007). Then, the sequences were computed with multiple sequences alignment methods by using ClustalW tools (Thompson, Gibson, & Higgins, 2002). The template was identified by using SWISS-MODEL server (Arnold, Bordoli, Kopp, & Schwede, 2006). The cyclic peptide ligands were constructed by using ACD/PhysChem Suite 12.0 with peptide sequence of PTLLFL and Ribavirin were applied as standards (Obayashi et al., 2008; Rowe et al., 2010; Spessard, 1998). Thus, the ligand-enzyme interactions were analyzed by using docking method and LigX tool of MOE 2008.10 software (Dias & de Azevedo, 2008; Tambunan, Parikesit, Prasetia, & Kerami, 2013).

**Results and Discussion**

Table 1, it shows the calculated free binding energy ($\Delta G_{binding}$) of flexible-ligand docking simulation from the best ligand. Free energy of binding for all the designed ligands was lower than the comparative ligand (ribavirin and peptide sequence of PTLLFL) (Guu, Dong, Wittung-Stafshede, & Tao, 2008). The ligand free energy of the peptide sequences CKTTC, CKKTC, PTLLFL and ribavirin were negative value, respectively, -20.9946, -18.0293, -14.4066, and -9.7573 kcal/mol. The negative and low value of $\Delta G_{binding}$ indicated the strong favorable bond between enzyme and ligand.

Table **1.** The ligand Free energy of binding and Ki results calculated using MOE 2008.10.

| Ligand | Estimated ΔG (kcal/mol) | Estimated Inhibition Constant/Ki (µM) |
|---|---|---|
| **Peptide Sequence of CKTTC** | -20.9946 | $1.9 \times 10^{-7}$ |
| **Peptide Sequence of CKKTC** | -18.0293 | $2.9 \times 10^{-4}$ |
| **Peptide Sequence of PTLLFL** | -14.4066 | $5.5 \times 10^{-1}$ |
| **Ribavirin** | -9.7573 | 0.52 |

These $\Delta G_{binding}$ values were parallel to the Ki values observed in Table 1. Ligand CKTTC and CKKTC showed the lowest Ki value, it could be estimated that the reaction equilibrium shifted to the complex formation. In general, all calculated Ki values were small (within micromolar range), indicating the formation of stable enzyme-ligand complexes. Every ligand showed reasonably low internal energy. It indicated that the docked conformation of the ligands were in their most favorable conformations (Zavodszky & Kuhn, 2005).

The substitution of Molecular Mechanics Generalized Born surface area (MM-GBSA) with the generalized Born (GB) approximate model of electrostatics in water was much more conducive for the PB electrostatics (Cui et al., 2008). It is already proven that MM-GBSA could effectively compute the binding free energy atomic/group contributions (Cui et al., 2008; Gohlke, Kiel, & Case, 2003). In this docking simulation, we utilized gas phase solvation (MacKerell et al., 1998).

In Table **2**, we presented residue contact of ligand peptide sequences of CKTTC, CKKTC, PTLLFL, and ribavirin. Red residue indicated amino acid residues which have highly important role of the interaction between the polymerase A and B1. Ligand CKTTC and CKKTC had residue contact with polymerase A (Gln 408, Glu 623) and polymerase B (Asp 2). The other side, residu contacts were occurred only in polymerase B1(Asp 2) for the ligand ribavirin and PTLLFL.

Table **2.** The residue contact of ligands in molecular docking simulation.

| Ligand | Residue contact | | | |
|---|---|---|---|---|
| | PA | *Score* (%) | PB1 | *Score* (%) |
| **Peptide Sequence of CKTTC** | **Gln 408** | 43.6 | **Asp 2** | 90.8 |
| | **Glu 623** | 64.0 | | |
| | Asn 708 | 45.3 | | |
| **Peptide Sequence of CKKTC** | **Gln 408** | 21.1 | **Asp 2** | 12.3 |
| | **Glu 623** | 49.5 | | 32.4 |
| | | | | 74.0 |
| **Peptide Sequence of PTLLFL** | Ser 405 | 24.7 | **Asp 2** | 77.6 |
| **Ribavirin** | Ser 709 | 18.1 | **Asp 2** | 36.4 |

Ligand CKTTC and CKKTC are good residue contact because Gln 408 plays an important role for PB1 binding. The Gln 408 and Asn 412 always form the hydrogen bonds with Asn 4 and Val 3 from PB1 respectively. At the same time, the residue 2-4 of PB1 tends to form β sheets and the regular hydrogen bonds between β sheets. During recognition process between PA and PB1, these interactions are important to the orientation of PB1 (H. Liu & Yao, 2010). The residue from 619-623 is very important for PA binding, which is in good agreement with in vitro binding assay results that the deletion of residues 619-630 for PA largely weakens the binding ability of PA to PB1 (Obayashi et al., 2008). If we compare all of the ligand, ligand CKTTC and CKKTC are the best ligand. Hopefully, this ligand can block interaction between PA and PB1.

In Figure **1**, we presented the ligand-receptor interaction diagrams. This is a depiction of relatively strong connections among hydrogen bonds as well as electrostatic or charge-transfer interaction between a ligand and the residue protein. The diagrams were elucidated by using LigX tool of MOE 2008.10 (Qamar et al., 2014).

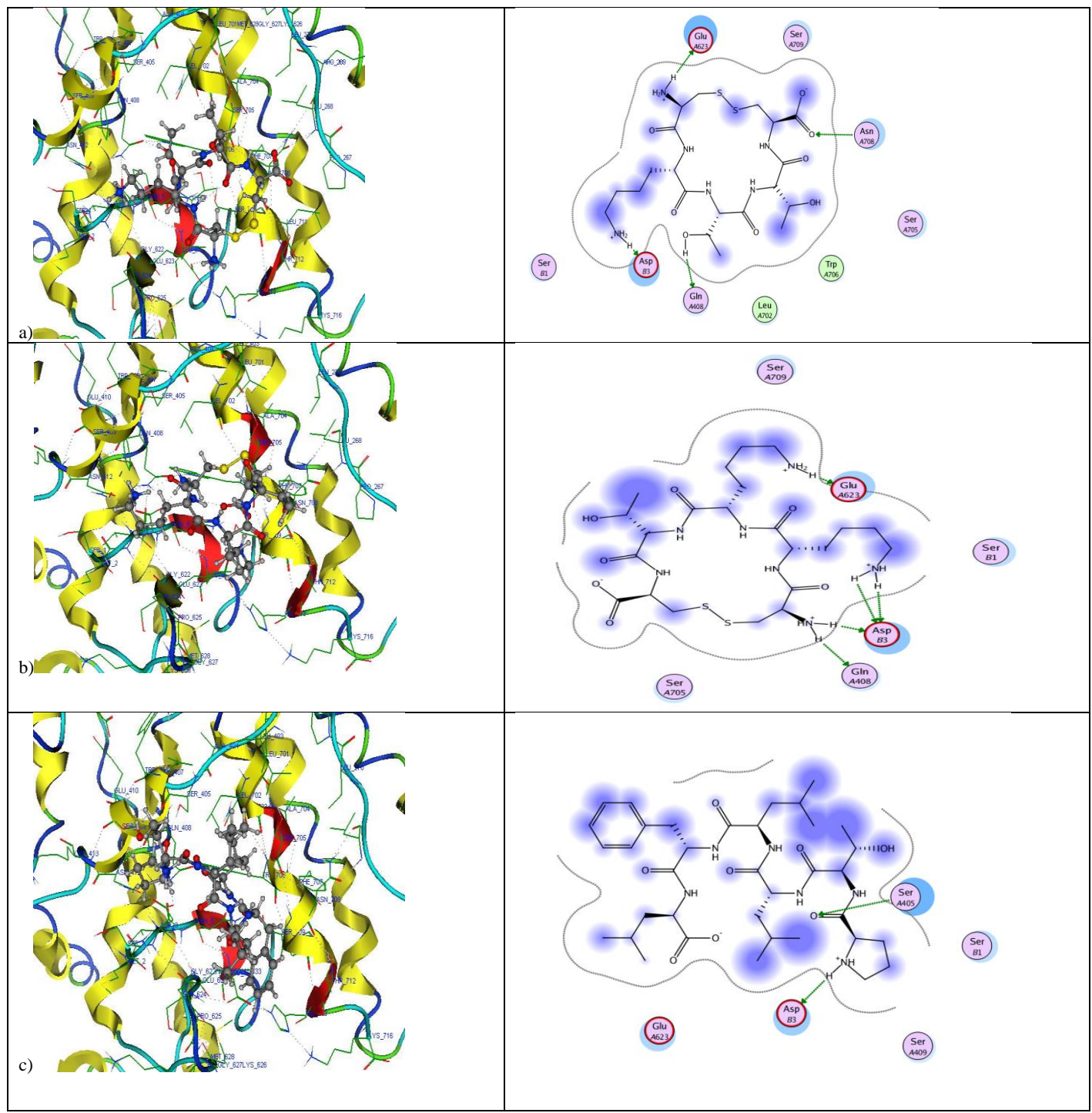

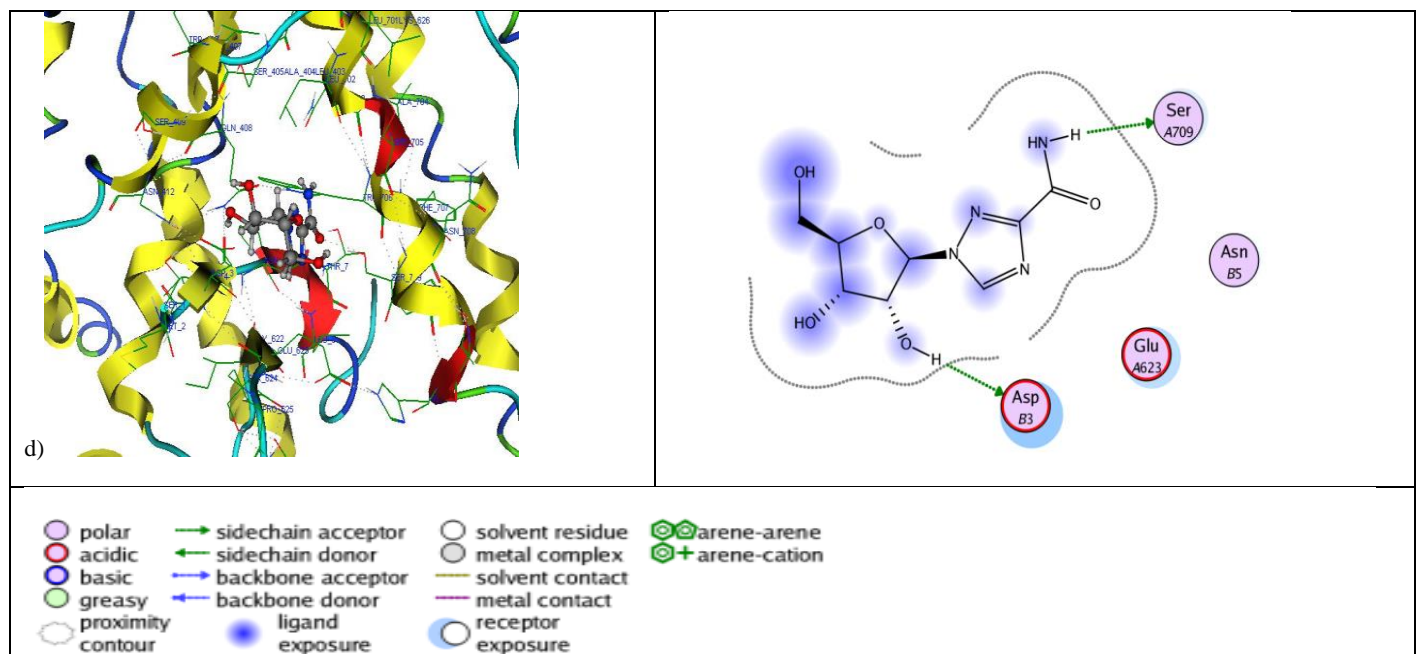

**Figure1.** Interaction plots illustrating between ligand and the respective amino acid residues. CKTTC (a), CKKTC (b), standard ligand PTLLFL (c), ribavirin (d).

## Conclusions

It is shown in the Figure 1 that the interaction complexity of the ligand-enzyme could be a contributing factor to develop peptide sequences of CKTTC and CKKTC as H1N1 drug candidates.

## Acknowledgments

The authors would like to thanks Hibah BOPTN Ditjen Dikti No 0970/H2.R12/HKP.05.00/2014 for providing the research grant. Usman Sumo Friend Tambunan supervised this research, Djati Kerami was giving important suggestion to improve our pipeline, Harry Noviardi was working on the technical details, while Arli Aditya Parikesit was writing the manuscript and annotated the data. Thanks also go to Niken Widiyanti for proof reading this manuscript.